# Rotation Effects on the Target-Volume Margin Determination

Qinghui Zhang

Department of Radiation Oncology, University of Nebraska Medical center, Omaha, Nebraska 68198

Department of Medical Physics, Memorial Sloan-Kettering Cancer Center, New York, NY 10065

Weijun Xiong, Maria F. Chan, Yulin Song, Chandra Burman

Department of Medical Physics, Memorial Sloan-Kettering Cancer Center, New York, NY 10065

**ABSTRACT**

Rotational setup errors are usually neglected in most clinical centers. An analytical formula is developed to determine the extra margin between clinical target volume (CTV) and planning target volume (PTV) to account for setup errors. The proposed formula corrects for both translational and rotational setup errors and then incorporated into margin determination for PTV.

## I. INTRODUCTION

It is well known that setup errors during treatment compromise the precision of radiation treatment. To improve the accuracy of patient positioning during treatment, On Board Imager (OBI) system has been developed and commonly used in present clinical centers[1]. Though six degrees of freedom (DOF) registration of cone-beam CT (CBCT) and planning CT can determine translational and rotational setup errors [2] in practice conventional couches do not allow rotational corrections. Therefore six DOF couch has been introduced to radiation clinics [3]. However, six DOF couches are not widely installed due to the various reasons including high cost.

In practice, most setup adjustments are applied to only translational shifts and rotational positioning errors are routinely disregarded throughout the patient's treatment. It is known that clinical target volume (CTV) to planning target volume (PTV) margin is determined by setup errors and possible motion during the treatment [4-10]. The margin should account for both translation and rotation setup uncertainties. The correction of translation setup will lead to a reduction of the margin and thereby spare more normal tissues. However, the oversight rotation error should result in a relatively larger margin. It was found that those rotation errors might cause dosimetric errors during clinical treatments [11]. Thus, it is interesting to derive an analytical formula to estimate the extra margin in the planning stage.

There are two purposes of this paper: (1) to derive the setup error distribution for a group of patients; and (2) to estimate the maximum setup error from rotations for each tumor size.

## II. METHODS

In this section we will: (1) derive the setup error distribution for a group of patients; (2) estimate the maximum setup error from rotation for each tumor. In our derivation, we only consider rigid-body registration.

For a patient setup, one must register the patient's CBCT images to the patient's planning CT images. For two rigid-body registrations, two things are needed for consideration: translation and rotation. In the following sections, we will discuss the registration of two brain tumors. The procedure we present here holds the truth when we register bones or other rigid structures.

Both tumors have N voxels in both the CT image and CBCT image. The CT image has a coordinate $\vec{x}_i (i=1,...,N)$ and CBCT image has another coordinate $\vec{y}_i (i=1,...,N)$. Without rotation, only translation is left for registration. Therefore, one tries to find the minimum difference between the tumors in the CBCT and CT images. In other words, we try to find a vector $\vec{A}$ such that:

$$I = \sum_{i=1}^{N} (\vec{x}_i - \vec{y}_i - \vec{A})^2 \tag{1}$$

is minimal. One can easily find that the solution of this equation is:

$$\vec{A} = \frac{1}{N}\sum_{i=1}^{N}\vec{x}_i - \frac{1}{N}\sum_{i=1}^{N}\vec{y}_i = \vec{x}_{CM} - \vec{y}_{CM} \tag{2}$$

Notice that we have used $\vec{x}_{CM} = \frac{1}{N}\sum_{i=1}^{N}\vec{x}_i$ and $\vec{y}_{CM} = \frac{1}{N}\sum_{i=1}^{N}\vec{y}_i$ to represent the center of mass (CM) of the tumors in the CT and CBCT images. Eq. (2) means that we translate the CM of tumor of CBCT to that of CT. However, rotation always exists so it needs to be included in our registration. Suppose the rotation matrix is R, then we try to match the CBCT image to the CT image by translation and rotation. Thus, we try to minimize:

$$I = \sum_{i=1}^{N}((\vec{x}_i - \vec{x}_{CM} - R(\vec{y}_i - \vec{y}_{CM}))^2 \tag{3}$$

One can determine the rotation matrix or R from Eq. (3). However, we will not discuss the method to determine R here. Instead, we will discuss the mismatch between those two tumors with this known rotation matrix. The above can be understood in the following way: For a rigid-body registration, one needs to first match the CM of the tumor (ROI) in the CT image with the CM of tumor in the CBCT image. After that, one can rotate the CBCT image to match the shape of tumor in the CT image. Using the traditional three Euler angles—roll ($\theta$), pitch ($\psi$), and yaw ($\phi$)—the rotation matrix can be expressed as (pitch-roll-yaw):

$$R = \begin{pmatrix} 1 & 0 & 0 \\ 0 & \cos\psi & \sin\psi \\ 0 & -\sin\psi & \cos\psi \end{pmatrix} \begin{pmatrix} \cos\theta & 0 & -\sin\theta \\ 0 & 1 & 0 \\ \sin\theta & 0 & \cos\theta \end{pmatrix} \begin{pmatrix} \cos\phi & \sin\phi & 0 \\ -\sin\phi & \cos\phi & 0 \\ 0 & 0 & 1 \end{pmatrix}. \tag{4}$$

It is well known that the final rotation matrix depends on the order of multiplication [12-13]. Using the first-order approximation, (i.e., $\sin(\theta) \approx \theta$ and $\cos\theta \approx 1$, with the same approximation for roll and yaw) and ignoring second-order terms and cross-product terms, we have:

$$R = \begin{pmatrix} 1 & \phi & -\theta \\ -\phi & 1 & \psi \\ \theta & -\psi & 1 \end{pmatrix} \tag{5}$$

Its inverse (ignoring the second-order terms) is:

$$R^{-1} = R^T = \begin{pmatrix} 1 & -\phi & \theta \\ \phi & 1 & -\psi \\ -\theta & \psi & 1 \end{pmatrix} \tag{6}$$

when the rotation setup error is around $2° = 0.0349$ radian, it can be easily verified that the error introduced by this approximation used above is of the order of 0.000007 for the sin(x) function and 0.0006 for the cos(x) function. Eq. (5) is a first order approximation to rotation matrix. It is independent of the order of the multiplications. We have calculated all rotation matrices for different order of multiplications. We have found that when pitch, roll, and yaw are <5°, the maximum difference between the approximation and exact solution is 0.0084. Therefore, the small angle approximation is sufficient and accurate enough for practical clinical applications. Nevertheless, we encourage our readers to check the rightness of this approximation for their own data.

Unfortunately, 6-DOF couches are not available in many clinical centers that most of us do not apply couch rotation in our clinical practices. Therefore, even if we can match the CM of the CBCT image to that of the CT image, we will still have residual setup errors left for tumor points. After matching the CM of the tumor in the CBCT image with the CM of the tumor in the CT image (e.g., the isocenter of the linac), the corresponding tumor points in the CBCT image are $\vec{y}_i = R^T \vec{x}_i$. Therefore, the residual setup error from rotation for point $\vec{x}_i$ can be defined as:

$$\vec{V}^R_{\vec{x}_i} = R^T \vec{x}_i - \vec{x}_i = \begin{pmatrix} 0 & -\phi & \theta \\ \phi & 0 & -\psi \\ -\theta & \psi & 0 \end{pmatrix} \vec{x}_i \qquad (7)$$

Here $\vec{V}^R_{\vec{x}_i} = (V^R_{\vec{x}_{i1}}, V^R_{\vec{x}_{i2}}, V^R_{\vec{x}_{i3}})^T$ is a vector. Ideally, we try to identify a CTV-PTV margin large enough to cover all those residual setup errors. Thus, we will try to find the maximum value of the residual setup errors for all points. This, of course, will depend on the shape of tumor and is specific to the patient. One can always use an ellipsoid function to fit each tumor:

$$\frac{x_{i1}^2}{a^2} + \frac{x_{i2}^2}{b^2} + \frac{x_{i3}^2}{c^2} = 1 \qquad (8).$$

Here (1,2,3) refers to the x,y and z components in the patient coordinate system. One can easily rewrite this equation by using the following format:

$$\begin{aligned} x_{i1} &= a\cos\beta\cos\lambda \\ x_{i2} &= b\cos\beta\sin\lambda \\ x_{i3} &= c\sin\beta \end{aligned} \qquad (9)$$

With

$$-\frac{\pi}{2} \leq \beta \leq \frac{\pi}{2} \qquad -\pi \leq \lambda \leq \pi$$

It is easily checked from Eq. (7) that:

$$\vec{V}_{\vec{x}_i}^R = \begin{pmatrix} -x_{i2}\phi + x_{i3}\theta \\ x_{i1}\phi - x_{i3}\psi \\ -x_{i1}\theta + x_{i2}\psi \end{pmatrix} \qquad (10)$$

Bringing Eq. (9) into Eq. (10), one has (the first row):

$$V_{\vec{x}_{i1}}^R = -x_{i2}\phi + x_{i3}\theta = -b\phi\cos\beta\sin\lambda + c\theta\sin\beta \qquad (11)$$

By taking $\frac{\partial V_{\vec{x}_{i1}}^R}{\partial \beta} = 0$ and $\frac{\partial V_{\vec{x}_{i1}}^R}{\partial \lambda} = 0$, we have following two equations:

$$b\phi\sin\beta\sin\lambda + c\theta\cos\beta = 0 \qquad (12)$$

and

$$b\phi\cos\beta\cos\lambda = 0 \qquad . \qquad (13)$$

Using the solutions of Eq. (12) and Eq. (13) and brought them into Eq. (11), we found that the maximum residual setup error for the x component is $\pm\sqrt{b^2\phi^2 + c^2\theta^2}$. Repeating that process for the y and z components, we have found that the maximum residual setup errors in the x-y-z components are:

$$\begin{aligned}\pm\sqrt{b^2\phi^2+c^2\theta^2} & \quad x \quad \text{component} \\ \pm\sqrt{a^2\phi^2+c^2\psi^2} & \quad y \quad \text{component} \\ \pm\sqrt{a^2\theta^2+b^2\psi^2} & \quad z \quad \text{component}\end{aligned} \quad (14)$$

Here the rotation angles are measured in radian. Therefore, as long as we can expand our CTV to include these maximum setup errors, we can cover the whole tumor. For typical setup errors $\theta \leq 1°, \phi \leq 1°$, and $\psi \leq 1°$, one can use Eq. (14) to calculate the maximum setup error for each tumor in each direction. For a stereotactic radiosurgery (SRS) case in which the tumor diameter is less than 4cm ($a \leq 2cm$, $b \leq 2cm$ and $c \leq 2cm$), one can easily estimate that the maximum residual setup error in each direction is 0.5 mm for the above setup error. If the setup error is approximately $2°$ in each direction, then the maximum residual setup error is approximately 1mm. If the tumor diameter is 10cm, then the setup error is 2.5mm. Therefore, the maximum setup error is proportional to the rotation angles and the tumor size. The probability distribution of this setup error can be expressed as:

$$P(\vec{V}_r^R) \propto \int \delta(|V_{r1}^R|-\sqrt{b^2\phi^2+c^2\theta^2})\delta(|V_{r2}^R|-\sqrt{a^2\phi^2+c^2\psi^2})\delta(|V_{r3}^R|-\sqrt{a^2\theta^2+b^2\psi^2})P(\theta,\phi,\psi)d\theta d\phi d\psi$$
(15)

In the Eq. (15), we have assumed that $P(\vec{V}_r^R)$ is a symmetry function for each component. The detail function format of $P(\vec{V}_r^R)$ depends on the function format of $P(\theta,\phi,\psi)$. $\theta, \phi$ are ranging from $-\pi$ to $\pi$. $\psi$ is ranging from $-\pi/2$ to $\pi/2$. The range of $P(\vec{V}_r^R)$ is [0; 1]. In the following, we will give a Gaussian approximation function to $P(\vec{V}_r^R)$. Under the assumption that $P(\vec{V}_r^R)$ is a symmetry function, one has:

$$\langle V_{r1}^R \rangle = \langle V_{r2}^R \rangle = \langle V_{r3}^R \rangle = 0 \tag{16}$$

and

$$\langle (V_{r1}^R)^2 \rangle = b^2 \langle \phi^2 \rangle + c^2 \langle \theta^2 \rangle, \quad \langle (V_{r2}^R)^2 \rangle = a^2 \langle \phi^2 \rangle + c^2 \langle \psi^2 \rangle, \quad \langle (V_{r3}^R)^2 \rangle = a^2 \langle \theta^2 \rangle + b^2 \langle \psi^2 \rangle \tag{17}$$

Here, $\langle (V_{r1}^R)^2 \rangle = \int (V_{r1}^R)^2 P(\vec{V}_r^R) d\vec{V}_r^R$ and similar definitions for other terms.

$\langle \phi^2 \rangle = \int \phi^2 P(\theta, \phi, \psi) d\phi$ is an average of $P(\theta, \phi, \psi)$ and similar definitions apply for the other two angles.

Therefore, to the second-order approximation, the residual setup error from the rotation is:

$$P(V_{r1}^R, V_{r2}^R, V_{r3}^R) \propto \exp(-\frac{(V_{r1}^R)^2}{2\langle (V_{r1}^R)^2 \rangle} - \frac{(V_{r2}^R)^2}{2\langle (V_{r2}^R)^2 \rangle} - \frac{(V_{r3}^R)^2}{2\langle (V_{r3}^R)^2 \rangle}) \tag{18}.$$

Eq. (18) is designed for considering rotation only. However, for clinical cases, we do not shift the CM of the tumor in a verification CBCT scan to that of the tumor in the planning CT. This is due to: (1) most treatment machines do not support the sub-millimeter couch shift. The couch step size is limited to 1 mm resolution; (2) no shift correction if the required shift is below the established action threshold; (3) for clinically large shifts, the post-correction residual errors between the CMs, although relatively small, are still non-zero owing to the reason explained in (1); and (4) potential patient motion during a prolonged setup and treatment. All of these, hereby, are termed as the residual translation errors.

Combining residual setup error from rotation and translation, we have the following setup errors:

$$P(V_{r1}, V_{r2}, V_{r3}) = \int P(\vec{V}_r^R) P(\vec{V}_r^T) \delta(\vec{V}_r - \vec{V}_r^T - \vec{V}_r^R) d\vec{V}_r^R d\vec{V}_r^T. \tag{19}$$

We assume the residual translation error as:

$$P(\vec{V}_r^T) = \frac{1}{(2\pi\sigma_{T1}^2)^{1/2}} \frac{1}{(2\pi\sigma_{T2}^2)^{1/2}} \frac{1}{(2\pi\sigma_{T3}^2)}$$
$$\exp(-\frac{(V_{r1}^T - V_{r01})^2}{2\sigma_{T_1}^2} - \frac{(V_{r2}^T - V_{r02})^2}{2\sigma_{T_2}^2} - \frac{(V_{r3}^T - V_{r03})^2}{2\sigma_{T_3}^2}) \quad . \tag{20}$$

Here, $V_{ri}^T$ ($i = 1,2,3$) represents three components from the residual translation setup error ($\vec{V}_r^T$), and $V_{r0i}$ ($i = 1,2,3$) represents three components of mean translation setup error, which is zero if the setup has perfect symmetry. However, it might not be zero, if the setup is not symmetrical. In that case, one obtains the total setup errors from translation and rotation (i.e., from Eq. (19)):

$$P(\vec{V}_r) =$$
$$\frac{1}{(2\pi\sigma_1^2)^{1/2}} \frac{1}{(2\pi\sigma_2^2)^{1/2}} \frac{1}{(2\pi\sigma_3^2)} \exp(-\frac{(V_{r1} - V_{r01})^2}{2\sigma_1^2} - \frac{(V_{r2} - V_{r02})^2}{2\sigma_2^2} - \frac{(V_{r3} - V_{r03})^2}{2\sigma_3^2}) \tag{21}$$

with

$$\sigma_i^2 = \sigma_{Ti}^2 + <(V_{ri}^R)^2> \quad i = 1,2,3. \tag{22}$$

It is clear that the total setup error becomes larger with this rotation (because $\sigma_i^2 \geq \sigma_{Ti}^2$ in Eq. 22). Here $\sigma_i$ is the component of the total setup error in the i direction. $\sigma_{Ti}$ and $\langle (V_{ri}^R)^2 \rangle$ are contributions from translation errors and rotation errors respectively.

The clinical usages of this derivation have three folds: (1) the measured rotation and translation setup errors can be used to calculate Eq. (21); (2) for a specific patient (with a known tumor size of a, b, or c), one can determine the setup error from the measured errors of rotation and translation; and (3) by using tumor size and Eq. (14), one can find critically acceptable rotation angles for a specific patient such that this patient's setup error is less than a given margin. This could be used to guide therapists in setting up patients.

Eq. (21) is size specific because it is associated with the specific size of tumors. For all sizes and all patients (with different tumor sizes), Eq. (15) changes to:

$$P(\vec{V}_r^R) \propto \int \delta(|V_{r1}^R| - \sqrt{b^2\phi^2 + c^2\theta^2})\delta(|V_{r2}^R| - \sqrt{a^2\phi^2 + c^2\psi^2})\delta(|V_{r3}^R| - \sqrt{a^2\theta^2 + b^2\psi^2})P(\theta,\phi,\psi)d\theta d\phi d\psi$$
$$P(a,b,c)da\,db\,dc \qquad (23)$$

Here P(a,b,c) is the probability distribution of the tumor sizes. Then, Eq. (17) changes to:

$$\langle (V_{r1}^R)^2 \rangle_{all} = \langle b^2 \rangle \langle \phi^2 \rangle + \langle c^2 \rangle \langle \theta^2 \rangle, \quad \langle (V_{r2}^R)^2 \rangle_{all} = \langle a^2 \rangle \langle \phi^2 \rangle + \langle c^2 \rangle \langle \psi^2 \rangle,$$
$$\langle (V_{r3}^R)^2 \rangle_{all} = \langle a^2 \rangle \langle \theta^2 \rangle + \langle b^2 \rangle \langle \psi^2 \rangle \qquad (24)$$

Here, $\langle a^2 \rangle = \int a^2 P(a,b,c)da\,db\,dc$, $\langle b^2 \rangle = \int b^2 P(a,b,c)da\,db\,dc$, and $\langle c^2 \rangle = \int c^2 P(a,b,c)da\,db\,dc$. The subscript means that this contribution from rotations are averaged over all patients. For all patients, the probability for the total setup error has the same function format as Eq. (21) but with different values of $\sigma_i$. The corresponding $\sigma_{i,all}$ for all patients and all sizes is

$$\sigma_{i,all}^2 = \sigma_{Ti}^2 + \langle (V_{ri}^R)^2 \rangle_{all} \qquad i = 1,2,3 \qquad (25)$$

III. RESULTS AND DISCUSSION

Currently, the frameless SRS at our institution is performed with an AKTINA Pinpoint radiosurgery System (AKTINA Medical, Congers, NY). The treatment is delivered on a Trilogy Linac (Varian Medical Systems, Palo Alto, CA). An AlignRT 3D optical surface imaging system (Vision RT, UK) is routinely used for both patient pre-setup and residual setup error measurements [14-15]. Our rotation and residual errors were measured with our AlignRT 3D surface imaging system, which has a much higher angular resolution than that of our CBCT imaging system because AlignRT uses a significantly large number of pixels for image registration. In our calculation, we limited our resolution to 0.1mm for translations and 0.1° for rotations. The accuracy of intermediate step calculations is up to 0.01mm.

In the tables I, diameters of 18 patients in the x, y, and z directions were measured which were used to determine a, b, c. The corresponding rotation angles were also measured. It was found that those rotation errors might induce dosimetric errors during clinical treatments when the rotation angle is large [11]. Thus, in our institution, when the data were measured, our clinical protocol requires that the maximum rotation angels should be ≤2° during our patient setup.

It is clear that the rotational margins needed for those rotation angles are very small except patient 18 which has a large tumor and two rotation angles are around 2 degree. But we need to point out here that for a SRS treatment, the tumor diameter is less than 4 cm. However this is not the case for other kind of treatments. The tumor volume can be very large.

The average rotation angles (yaw, roll, pitch) are -0.0°, 0.3°, and 0.3° and the corresponding standard deviations are 1.0°, 0.5° and 0.7°, respectively. The average values of the residual setup error from rotation in the three components of our data are 0.2 mm, 0.2 mm, and 0.1 mm, respectively. The standard deviations are 0.2 mm, 0.2 mm, and 0.2 mm, respectively.

Table I: The measured results of the 18 patients. The maximum residual setup error in the x, y and z components from rotations (denotes as $M_i$ ($i=1,2,3$)) are calculated according to Eq. 14.

| Patient | a (mm) | b (mm) | c (mm) | $\phi(^o)$ | $\theta(^o)$ | $\psi(^o)$ | $M_1$ (mm) | $M_2$ (mm) | $M_3$ (mm) |
|---|---|---|---|---|---|---|---|---|---|
| 1 | 7.0 | 7.2 | 7.3 | 0.6 | 0.5 | 0.9 | 0.10 | 0.14 | 0.13 |
| 2 | 7.5 | 6.2 | 5.2 | 0.4 | -0.2 | 0.3 | 0.05 | 0.06 | 0.04 |
| 3 | 10.8 | 11.0 | 10.8 | -0.7 | -0.2 | 0.1 | 0.14 | 0.13 | 0.04 |
| 4 | 6.3 | 5.7 | 7.8 | 0.3 | 0.8 | 1.0 | 0.11 | 0.14 | 0.13 |
| 5 | 8.3 | 8.3 | 7.7 | 0.4 | 0.2 | 0.7 | 0.06 | 0.11 | 0.11 |
| 6 | 6.5 | 7.1 | 7.0 | 0.4 | 0.2 | 0.3 | 0.06 | 0.06 | 0.04 |
| 7 | 8.4 | 5.1 | 7.7 | 0.1 | 0.3 | 0.1 | 0.04 | 0.02 | 0.04 |
| 8 | 7.5 | 7.6 | 7.2 | -0.7 | 0.6 | 0.5 | 0.12 | 0.11 | 0.10 |
| 9 | 10.2 | 11.3 | 8.7 | -0.7 | 0.6 | 0.1 | 0.17 | 0.13 | 0.11 |
| 10 | 9.2 | 10.5 | 10.1 | -1.0 | 0.5 | 0.6 | 0.20 | 0.19 | 0.14 |
| 11 | 4.9 | 6.2 | 5.5 | -0.1 | 0.2 | -0.6 | 0.02 | 0.06 | 0.07 |
| 12 | 8.2 | 8.2 | 7.4 | 0.7 | 0.5 | -0.6 | 0.12 | 0.13 | 0.11 |
| 13 | 7.7 | 8.5 | 8.4 | 1.1 | 0.8 | -0.7 | 0.20 | 0.18 | 0.15 |
| 14 | 8.8 | 9.9 | 10.0 | 0.8 | -0.1 | 0.1 | 0.14 | 0.12 | 0.02 |
| 15 | 5.6 | 6.2 | 5.3 | -0.7 | 0.3 | 0.4 | 0.08 | 0.08 | 0.05 |
| 16 | 8.2 | 8.2 | 8.1 | -2.2 | -0.6 | 1.3 | 0.33 | 0.36 | 0.21 |
| 17 | 8.6 | 10.9 | 9.1 | -1.2 | 0.2 | -0.4 | 0.23 | 0.19 | 0.08 |
| 18 | 19.0 | 16.0 | 21.0 | 2.3 | 1.5 | 2.0 | 0.85 | 1.06 | 0.75 |

We will not list the residual translational setup error for each patient. Instead the stand deviation of the translational setup error for each components is given here: $\sigma_{T1} = 0.53 mm$, $\sigma_{T2} = 0.65 mm$ and $\sigma_{T3} = 0.84 mm$. From the table I, one obtains $\langle a^2 \rangle \approx 80.54 \, mm^2$, $\langle b^2 \rangle \approx 80.25 \, mm^2$, $\langle c^2 \rangle \approx 84.89 \, mm^2$, $\langle \theta^2 \rangle \approx 0.0001$, $\langle \phi^2 \rangle \approx 0.0003$ and $\langle \psi^2 \rangle \approx 0.0002$. According to Eq. (24), we have $\langle (V_{r1}^R)^2 \rangle_{all} \approx 0.03\,mm^2$, $\langle (V_{r2}^R)^2 \rangle_{all} \approx 0.04\,mm^2$, and $\langle (V_{r3}^R)^2 \rangle_{all} \approx 0.02\,mm^2$. Thus the total set up errors for all patients and all tumor sizes are (using Eq. 25) $\sigma_{1,all} = 0.6mm$, $\sigma_{2,all} = 0.7mm$ and $\sigma_{3,all} = 0.9mm$. It is clearly that the difference between $\sigma_{i,all}$ and $\sigma_{Ti}$ is small. That is easily understood from table I where the average tumor volume and rotation are small. For a specific tumor size (Patient 18, for example), one has (using Eq. 17) $\langle (V_{r1}^R)^2 \rangle \approx 0.72\,mm^2$, $\langle (V_{r2}^R)^2 \rangle \approx 1.12\,mm^2$, and $\langle (V_{r3}^R)^2 \rangle \approx 0.56\,mm^2$. Then the corresponding total setup errors for this specific size are (using Eq. 22) $\sigma_1 = 1.0mm$ $\sigma_2 = 1.2\,mm$ and $\sigma_3 = 1.1\,mm$. Therefore, the contribution from rotation error cannot be ignored when the tumor size is around the 4cm even when the rotation error is still very small.

From the total setup error and systematic error, one can use the margin recipes of [5] to calculate the CTV-PTV margins for multi-fraction treatments and the margin recipes of [7-8] for single fraction treatments. If we ignore other systematic error except the non-coincidence between CBCT image isocenter and linac isocenter[16-20], the margins for three independent one-dimensional expansion such that 95% patients received the prescribed dose [8] can be approximated as

$$M_{i,all} = W_{0i} + 2.33\sigma_{i,all}$$

We assume that the isocenter differences between CBCT and linac is around 0.5mm in all three directions. The isocenter difference is machine-dependent. It also depends on the measurement methods that the physicists used. Nevertheless, it is always of the same order. Thus, the corresponding margins are $M_{1,all} = 1.9mm$, $M_{2,all} = 2.1mm$ and $M_{3,all} = 2.6mm$. Therefore, the margin for the single fraction treatment is around 2 to 3mm.

In this note, our method was demonstrated with SRS cases. We will continue this study for other disease sites where tumor volumes may be large and the rotation effects may be dosimetrically significant. In addition, the registration uncertainty from rotation also affects the residual setup error, this, too, will be further investigated in our future research work.

## IV. CONCLUSIONS

Rotational setup errors always exist in setup process during treatments. However, this kind of errors is often neglected in almost all clinics. To account for this omission, an extra margin needed for the margin between CTV and PTV. An analytical formula has been proposed which can be used to estimate this extra margin. A total setup error distribution is derived which include rotation effects. This distribution can be used for margin determination.